# Impact of MBE-grown (In,Ga)As/GaAs metamorphic buffers on excitonic and optical properties of single quantum dots with single-photon emission tuned to the telecom range

Paweł Wyborski,[1,*] Michał Gawełczyk,[2] Paweł Podemski,[1] Piotr Andrzej Wroński,[3] Mirosława Pawlyta,[4] Sandeep Gorantla,[5] Fauzia Jabeen,[3,6] Sven Höfling,[3,6] and Grzegorz Sęk[1]

[1]*Department of Experimental Physics, Wrocław University of Science and Technology, Wybrzeże Wyspiańskiego 27, 50-370 Wrocław, Poland*
[2]*Institute of Theoretical Physics, Wrocław University of Science and Technology, Wybrzeże Wyspiańskiego 27, 50-370 Wrocław, Poland*
[3]*Technische Physik, University of Würzburg and Wilhelm-Conrad-Röntgen-Research Center for Complex Material Systems, Am Hubland, D-97074 Würzburg, Germany*
[4]*Materials Research Laboratory, Faculty of Mechanical Engineering, Silesian University of Technology, Konarskiego 18A, Gliwice, 44-100, Poland*
[5]*Łukasiewicz Research Network – PORT Polish Center for Technology Development, ul. Stabłowicka 147, 54-066 Wrocław, Poland*
[6]*Faculty of Engineering and Physical Sciences, University of Southampton, SO17 1BJ Southampton, United Kingdom*

Tuning GaAs-based quantum emitters to telecom wavelengths makes it possible to use the existing mature technology for applications in, e.g., long-haul ultra-secure communication in the fiber networks. A promising method re-developed recently is to use a metamorphic InGaAs buffer that redshifts the emission by reducing strain. However, the impact of such a buffer causes also a simultaneous modification of other quantum dot properties. Knowledge of these effects is crucial for actual implementations of QD-based non-classical light sources for quantum communication schemes. Here, we thoroughly study single GaAs-based quantum dots grown by molecular-beam epitaxy on specially designed, digital-alloy InGaAs metamorphic buffers. With a set of structures varying in the buffer indium content and providing quantum dot emission through the telecom spectral range up to 1.6 μm, we analyze the impact of the buffer and its composition on QD structural and optical properties. We identify the mechanisms of quantum dot emission shift with varying buffer composition. We also look into the charge trapping processes and compare excitonic properties for different growth conditions with single-dot emission successfully shifted to both, the second and the third telecom windows.

## I. INTRODUCTION

Non-classical emitters based on semiconductor quantum dots (QDs) fabricated in epitaxial technology have been demonstrated as promising candidates for fundamental components of many nanophotonic devices, including lasers [1–3], optical amplifiers [4,5] and broadband sources [6–8]. Especially in quantum information processing, QDs have been proven as efficient single-photon sources (SPS) to also realize ultra-secure quantum communication [9–12]. As perhaps the most crucial property of QD-based structures, emission in the range of highly transmissive second and third telecom windows [9,11,12] has been demonstrated, enabling integration with the existing silica-based optical-fiber infrastructure for the implementation of long-haul telecommunication. Moreover, on-demand QD-based high-purity emission of single, and optionally indistinguishable photons, and generation of polarization-entangled photon pairs have been shown [13]. Furthermore, many crucial milestones for applying QD devices in secure quantum information technology have been achieved. These include demonstrating advanced nanophotonic devices with high efficiency of single-photon emission [10,14–16], operation at elevated temperatures [12,17,18], integration with silicon platform [19], and the deterministic fabrication of QD structures [14].

---

[*]Corresponding author (pawel.wyborski@pwr.edu.pl)

The selection of the material system and the control of the QD parameters through the band structure engineering, including the mismatch of materials lattice constants and the resulting strain field, allows controlling the emission energy and adjusting it to the selected spectral range, particularly second and third telecom windows [20]. The first choice of materials for the telecom windows would be InAs QDs grown on InP substrate. They have already been demonstrated as SPSs in this range [11,12,15,21–28], with several implementations of efficient single photon emitters, additionally generating indistinguishable photons [29–31] and entangled photon pairs (also at elevated temperatures) [27,32,33], and realizations of quantum teleportation [29,33].

InAs/GaAs QDs typically emit below 1200 nm [9,20], and their tuning to the telecom range is more demanding, mainly due to a significantly larger lattice mismatch and built-in strain. But the technology itself has advantages over the InP-based one: it is more mature in some aspects and allows growing highly reflective distributed Bragg reflectors (DBRs), which is more challenging on the InP substrates due to lower refractive index contrast of the lattice-matched semiconductors. Several approaches allow to achieve the telecom spectral range with GaAs-based structures, like confinement and strain engineering [34–36], modification of the QD composition or QD size increase [37–39], or growing the dots on a metamorphic buffer layer (MBL) made of InGaAs [20,40–45]. However, so far, only the latter led to the demonstration of single-photon emission in the third telecom window characterized by the lowest signal attenuation in the fibers [20,41,46]. This solution is promising for applications and allows for easy fabrication of QD emitters inside photonic structures [47,48], and integration with the GaAs-based DBRs [20]. In that context, the major milestone demonstrations concern the MOVPE-grown QD systems: emission of polarization-entangled photon pairs [49,50] and indistinguishable photons (also with resonant excitation and in on-demand mode) [51–53], and the implementation of a high-performance photonic structure based on a circular Bragg grating [47,53]. Further, piezo tuning of QD emission energy [54,55] and QD-based coherent control of spin-qubit [56] have also been shown. In contrast, for the MBE-grown systems, emission energy shifts towards telecom for both laser structures [42,57] and single-QD structures [40,43,45,46,58] have only been reported. Therefore, this work attempts to fill this gap by performing a systematic and comprehensive study, because there is still very limited knowledge on the single-dot properties of such GaAs-based telecom-wavelength QDs grown by MBE on an MBL, especially in the wide range of emission energies. Since the QDs grown by MBE can significantly differ from their MOVPE counterparts they should be carefully investigated to fully verify their potential as non-classical radiation sources at the telecom and exploring them promises to overcome at least some of the limitations met and reported for MBL-based QDs grown in MOVPE.

Implementing efficient SPS for real applications in secure quantum communication has a few requirements, e.g., high source efficiency, high generation rate and purity of single-photon emission, long coherence of emission, and operation around room temperatures (or simple cooling method). A significant part of these is directly related to single QD properties [9]. Growing QDs on an MBL, besides the emission energy change, enables modifying other application-relevant properties such as emission polarization, carrier lifetimes and carrier loss processes, the structure of confined states, and properties of excitonic complexes like binding energy and fine structure splitting (FSS). Modifying the strain conditions during growth may change the structural QD parameters (size and indium content) [37] and can also influence the charge environment around the QDs [51], thus affecting all the QD optical properties.

Here, we analyze the structural and optical properties of single InGaAs QDs for several structures with variable indium content in the MBL resulting in emission wavelengths from 1.1 μm to 1.6 μm. We use a combination of several complementary spectroscopic techniques, atomic force microscopy (AFM), scanning electron microscopy (SEM), and high angle annular dark field scanning transmission electron microscopy (HAADF-STEM), all supported by calculations within the eight-band **k·p** theory to confront the electronic structure with the QD structural and optical properties, including charge environment and carrier escape paths. We trace all the QD optical parameters for different MBL compositions, indicating the importance of the MBL influence on QD properties going beyond the simple emission energy shift. We show that tuning the emission towards the telecom range with increasing MBL In concentration is accompanied by a slight increase of the photoluminescence (PL) linewidth and almost MBL composition-independent polarization anisotropy of emission. Additionally, we find a slight increase of FSS and a mostly constant value of biexciton binding energy as a function of the emission energy. Moreover, we observe charge environment changes and their influence on emission properties and PL decay dynamics. Eventually, we show good single-photon emission characteristics in the second and third telecom windows regardless of the emission shift.

## II. METHODS AND MATERIALS

Atomic force microscopy (AFM) measurements were performed using a commercial "Anfatach Level AFM" device, offering a spatial resolution of about 2 nm (in-plane) and a scanning resolution of about 0.2 nm. An intermittent sample contact mode was used with a cantilever having a resonance frequency of about 210 kHz. For scanning electron microscopy (SEM) a "Hitachi S5000 HRSEM" microscope was used, offering electron beam acceleration of up to 30 kV. Detection of secondary electrons allowed for surface topography measurements with a maximum resolution of 0.6 nm. Measurements of buried QDs were carried out using HAADF-STEM imaging based on the signals of elastically scattered electrons in the high-angle collection range of 79.5 – 200 mrad, ensuring the observation of atomic (~ $Z^2$) contrast in the studied structures. The "S/TEM TITAN cube G2 80-300" microscope by TFS (Thermo Fisher Scientific) was used, equipped with a high brightness field emission gun (X-FEG) electron source, double Cs correctors ensuring spatial resolution of ~ 70 pm, and operated with an accelerating voltage of 300 kV. Moreover, the super-X 4-detector X-ray energy dispersive spectrometer (EDS) on it allowed us to characterize the composition of the sample. Samples for STEM characterization were prepared using the FIB method (employing the "FEI SEM/XE-PFIB" and "FEI SEM/GA-FIB" microscopes).

Photoluminescence measurements were performed using a standard experimental setup, offering a high-spatial-resolution option based on a microscopic objective (NA = 0.4) focusing the excitation laser and collecting the PL signal with a spatial resolution below 2 µm. Liquid-helium continuous-flow cryostat was used for obtaining low temperatures (down to about 5 K) to reduce thermally-activated carrier losses and the influence of acoustic phonons on the spectral lines' broadening. It also allowed for temperature-dependent studies. Non-resonant continuous-wave (cw) excitation of the sample was performed with a 640 nm laser line. For quasi-resonant excitation and PL excitation (PLE) measurements, a tunable cw laser with a special setup for laser-line filtering was used [59]. Detection based on a 1-m-focal-length spectrometer coupled to a liquid-nitrogen-cooled InGaAs linear array detector with an effective spectral resolution of 20 µeV was used for both PL and PLE. Time-resolved photoluminescence (TRPL) characterization and photon statistics measurements using Hanbury Brown and Twiss configuration in time-correlated single-photon counting mode were also performed. For this, a 0.32-m-focal-length monochromator for spectral filtering with detection by fiber-coupled NbN superconducting nanowire single-photon detectors, connected to a multichannel picosecond event-timer with an overall temporal resolution of the experimental setup of about 80 ps, were used. For pulsed excitation in TRPL, an 805 nm semiconductor diode laser with pulse-train control possibility (2.5-80 MHz) and approximately 50-ps-long pulses was employed.

Theoretical QD modeling was performed based on the available structural data. Initially, a lens-shaped geometry with a height of 5 nm and a base diameter of 40 nm, placed on a 1.2-nm-thick wetting layer, was assumed. A slight 20% in-plane asymmetry was introduced to make the model not overly idealized. Initially, a homogeneous material composition was assumed in a QD. Then, Gaussian averaging (σ = 0.9 nm) of the material distribution was performed to simulate the unavoidable interdiffusion of atoms at interfaces. There were no premises in the structural data for a significant and systematic QD geometry change between the samples. Therefore, constant geometry was kept, and only modification of the composition of the MBL layer and QDs between the modeled structures was considered. For the dots modeled this way, the strain field within the continuum elasticity theory and the shear-strain-induced piezoelectric field up to the second order in the strain-tensor elements were calculated. The eigenstates of electrons and holes were calculated using an implementation [60] of the **k·p** method [61] that includes the effects of strain, piezoelectric field, and spin-orbit coupling. The explicit form of the Hamiltonian used can be found in Ref. [62], and the material parameters in Ref. [63] and references therein. Then, states of excitons and higher-order carrier complexes were calculated within the configuration-interaction method with the basis of 32 electron and 32 hole single-particle states. Finally, the optical transition dipole moment and the resulting emission polarization and recombination times [64] were derived in the dipole approximation [65].

The investigated structures were grown by MBE on GaAs (001) substrate starting with a GaAs buffer layer. Then, graded-composition InGaAs MBL was grown with different maximal indium content in the top part to obtain the QD emission redshift. On top of the MBL, InAs material was deposited to form QDs. In the final step, QDs were capped with an InGaAs layer. Due to natural atom diffusion processes, the obtained QDs are effectively composed of InGaAs. Figure 1(a) presents a layer structure scheme of the investigated samples, whereas Table I lists all the studied structures, differing in the MBL composition and thickness. More insight into the sample growth can be found in Ref. [40] for structures with indium content in the top part of MBL: 0.15, 0.20, 0.24, 0.29, and in Ref. [46] for the structure with 0.38 indium content. For the In-38% structure, additional five GaAs/AlAs DBR pairs were grown. On top of each structure, a reference (optically inactive) QD layer was grown for the structural characterization. For systematic and repeatable characterization of single QD emission, a combination of electron-beam lithography and wet etching was performed to obtain cylindrically shaped mesas for all structures except In-38%. For the In-38% structure, circular apertures in a thin silver layer deposited on the sample surface were made, offering similarly repeatable characterization without etching necessity, taking advantage of the low spatial (and hence spectral) density of these QDs [46].

TABLE I. Structure parameters: indium content and the measured thickness of the MBL layer.

| Sample | In content at the top of MBL (%) | MBL thickness (nm) |
|---|---|---|
| In-15% | 15 | 780 |
| In-20% | 20 | 960 |
| In-24% | 24 | 1140 |
| In-29% | 29 | 1140 |
| In-38% | 38 | 1200 |

However, one needs to keep in mind that uncapped QDs can differ from buried dots, even significantly [66]. It is known that capping can impact material intermixing, additional material accumulation, and cause strain changes. Therefore, the structural properties of the top-layer dots give just an approximate insight into the morphology and density. Nonetheless, one can perform less demanding characterization by atomic force microscopy (AFM) and scanning electron microscopy (SEM) to still get some information. Figures 1(b) and 1(c) present examples of SEM and AFM images of QDs for the In-29% structure. SEM-based results show an average QD height of about 3 nm and a diameter of about 25 nm. The AFM characterization of QDs' shape shows only slight in-plane asymmetry with an average lateral aspect ratio (LAR) of about 1.2. High-resolution STEM characterization allowed finding parameters of buried optically active QDs (width, height, and chemical composition of QDs and MBL material), which we consider more realistic (but with some other limitations). For the In-29% structure, an exemplary lens-shaped QD cross-section is presented in Fig. 1(d). Statistical analysis of many QDs provides average QD parameters: about 5 nm ($4.71 \pm 0.25$ nm) height ($H$) and 40 nm ($37.2 \pm 2.0$ nm) diameter ($D$). The values are higher than those from SEM characterization, showing the expected influence of the capping layer on the final QD dimensions (and most likely composition). Due to the limitations of the STEM lamella fabrication accuracy (final QD cross-section location) and the following analysis of images, it is still possible that the size of QDs can be slightly underestimated in this case, mainly for the QD height. The chemical composition distribution inside the dot is also essential, especially for indium with a tendency to segregate, and can significantly modify the QD optical properties. The EDX spectroscopy characterization accuracy of such a small QD composition is inherently limited. Thus, indium content inside QDs was obtained by comparing the emission energy with our simulation. However, the EDX characterization shows the indium composition of MBL and the cap layer (about 2/3 of the top MBL value) is consistent with the optically measured values. Figure 1(e) shows a line scan of the electron beam parallel to the growth direction and crossing the QD layer for the In-29% structure.

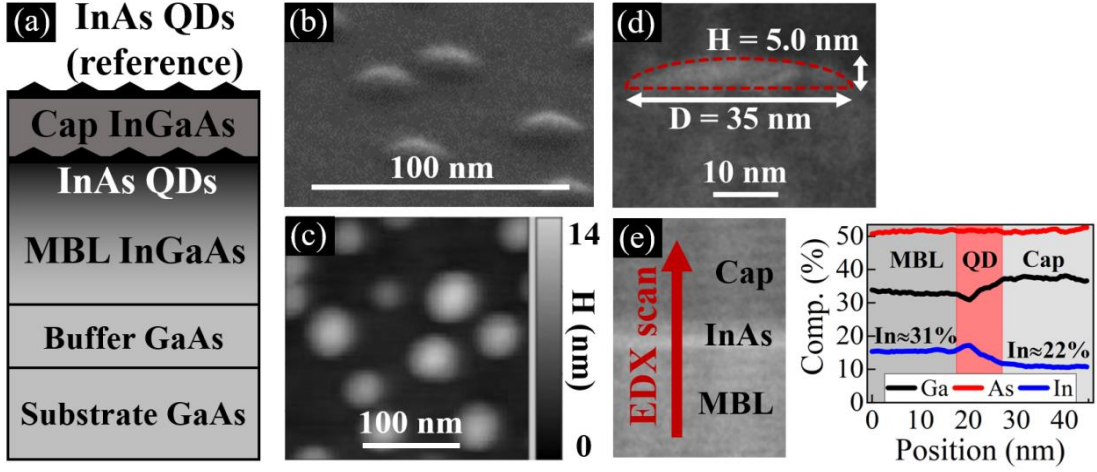

FIG. 1. (a) Scheme of the investigated structure with InGaAs QDs. (b) SEM image of QDs for In-29% structure. (c) AFM image of QDs for In-29% structure. (d) Cross-section HAADF STEM image of QD for In-29% structure. (e) EDX linescan characterization of In-29% structure.

## III. RESULTS AND DISCUSSION

### A. MBL impact on the emission properties of InGaAs QDs

The composition of the top part of the MBL directly impacts the QD emission energy, shifting it to the second (In-29% structure) and the third telecom windows (In-38% structure), as shown in Fig. 2(a) with low-temperature PL spectra for QDs ensemble. This results from the reduced lattice mismatch between the MBL and InAs QD layer, and thus decreased compressive strain in QDs compared to pure InAs/GaAs system. Strain reduction directly shifts QD energy levels. It can also lead to the growth of larger QDs. Both these effects have a direct influence on the emission energy. Another factor directly influencing the confined states' energy is the composition of quantum dots. The increased indium content at the top of the MBL (located directly underneath the QD layer) can favor a larger concentration of indium inside the dots, which also shifts the emission to longer wavelengths. Based on the STEM characterization (see Fig. 1(d) for the In-29% structure; similar for other samples not shown here), the average size of QDs does not change significantly when the composition of the MBL is changed, emphasizing the importance of the indium content modification. These observations are consistent with the results of our theoretical calculations confirming the strain modification (driven by the MBL composition change) and indium content inside QDs as sources of shifting the QD emission energy.

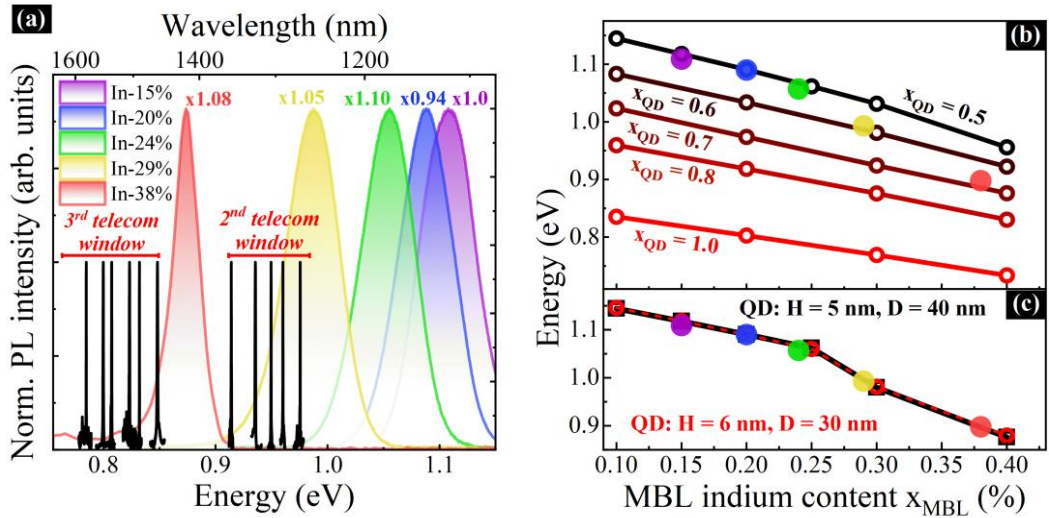

FIG. 2. (a) Normalized PL spectra for QDs ensemble for structures from 15% to 38% indium content in the top part of MBL (temperature: 10 K, excitation wavelength: 532 nm, cw, multipliers with respect to PL intensity of the In-15% sample) with overlaid single QD emission lines in the range of second and third telecom windows (temperature: 5 K, excitation wavelength: 640 nm, cw). (b) Calculated QD emission energies as a function of MBL indium content ($x_{MBL}$) for varying QD indium content ($x_{QD}$: 0.5, 0.6, 0.7, 0.8, 1.0) and optically-measured emission energies for the investigated structures. (c) Results including QD indium content change (from 0.50 to 0.65) for two QD geometries: $H = 5$ nm, $D = 40$ nm, and $H = 6$ nm, $D = 30$ nm (solid black and dashed red lines).

Figure 2(b) shows the calculated exciton ground-state energies (lines) together with energies of the PL maxima for the ensemble of dots as from Fig. 2(a) (full symbols) of the examined structures as a function of the In content in the MBL layer ($x_{MBL}$). Each line corresponds to changing $x_{MBL}$ only with fixed geometry and In content inside a QD ($x_{QD}$). The line for $x_{QD} = 0.5$ corresponds very well with the first three experimental points. However, the last two points ($x_{MBL} = 0.29$ and 0.38) do not follow this trend. We assume a higher $x_{QD}$ for $x_{MBL} > 0.25$ to reproduce this result, with linearly varying $x_{QD}$ from $x_{QD} = 0.6$ at $x_{MBL} = 0.3$ to 0.7 at $x_{MBL} = 0.4$ (see black line in Fig. 2(c)). The additional red line in Fig. 2(c) presents the results obtained in analogous simulations but for a slightly modified QD geometry: $H = 6$ nm, $D = 30$ nm, and $x_{QD}$ slightly increased by 0.02 compared to the calculations discussed before. The result almost coincides with the calculations for the nominal values, showing the negligible influence of the QD geometry. These modified dimensions are within the uncertainty limits of determining the QD geometry from the structural data bu also justified due to ensemble inhomogeneity. We observe a broadening of the PL peaks of about 50–55 meV for structures In-15%, In-20%, In-24%, and In-29% and approx. 25 meV for In-38%, originating from the distribution of quantum dot parameters (size or dot composition) within an ensemble and typical for InAs QDs [39,67-69], with a possible additional contribution due to spatial alloy fluctuations causing local band gap and strain changes [70].

Figure 2(a) shows also normalized single QD emission lines within the telecom windows, confirming the possibility of applying these structures in long-distance quantum communication schemes in the fiber networks. The possibility of observing individual QDs, even for structures with the largest indium content in the MBL layer, confirms the good optical quality of our structures. The ensemble emission provides only averaged information about the dots. For applications, especially in nanophotonics and non-classical light sources for quantum technologies, single QDs are of particular interest. Thus, we focus in this work on single QD properties. Figure 3(a) shows the recorded PL linewidths (full width at half maximum - FWHM) for many studied single QDs from samples with different MBL compositions. Error bars in Fig. 3(a) and all the following figures correspond to the standard deviation obtained from the fitting procedure. Regardless of the emission energy, we observe relatively large FWHM values significantly above 100 μeV, however comparable to InAs/InP QDashes [18,71] and low-strain QDs [72,73]. The linewidths are most likely related to the local electric environment fluctuations (due to discharging and recharging charge traps near QDs) resulting in significant spectral diffusion, i.e. fast fluctuations of QD emission energy broadening the time-integrated spectral lines [74]. This suggests the existence of built-in charge traps in the vicinity of the dots. The median FWHM value changes from 170 μeV for 1.1 eV emission to 300 μeV at 0.8 eV, showing a slight dependence on the MBL composition and indicating an increase in the concentration of defects in the vicinity of QDs with MBL indium content. Achieving linewidths significantly below 50 μeV, comparable to the best for InAs/GaAs QDs [41,75-77], will require reducing the number of defects near QDs by optimizing the growth parameters and MBL composition gradient or by adjusting the composition of the capping layer [45,78]. It is also possible that interaction of the excitons confined in the dots with charge traps is enhanced for structures with weaker confinement potential, i.e., those with lower composition difference between MBL and QDs - high In content samples emitting at longer wavelengths. For those, larger in-plane elongation can be expected (see the discussion in the next subsections), which can increase this cross-section even further. In addition, surface processing (formation of mesas or aperture structures) can also contribute to the spectral diffusion and the inhomogenous linewidth. Both can provide additional charge traps: (i) for mesas, due to the surface states on the mesa sidewalls which are in a direct vicinity of some of the dots; (ii) for metallic masks, due to possible charge traps on the semiconductor-metal interface, which, however, are separated from QDs by the cap thickness. Therefore, the impact of mesas is usually a bit more pronounced, especially for smaller mesas (which we have verified for another sample with QDs on MBL). Please note, that we observed linewidths above 70 μeV for all emission lines (FIG. 3(a)), indicating a significant degree of charge fluctuations with a trend to increase for smaller emission energies, both regardless of the type of structure processing (metallic mask or mesa). Thus, the broadening must be at least partly (or even predominantly) related to the quality of the MBL.

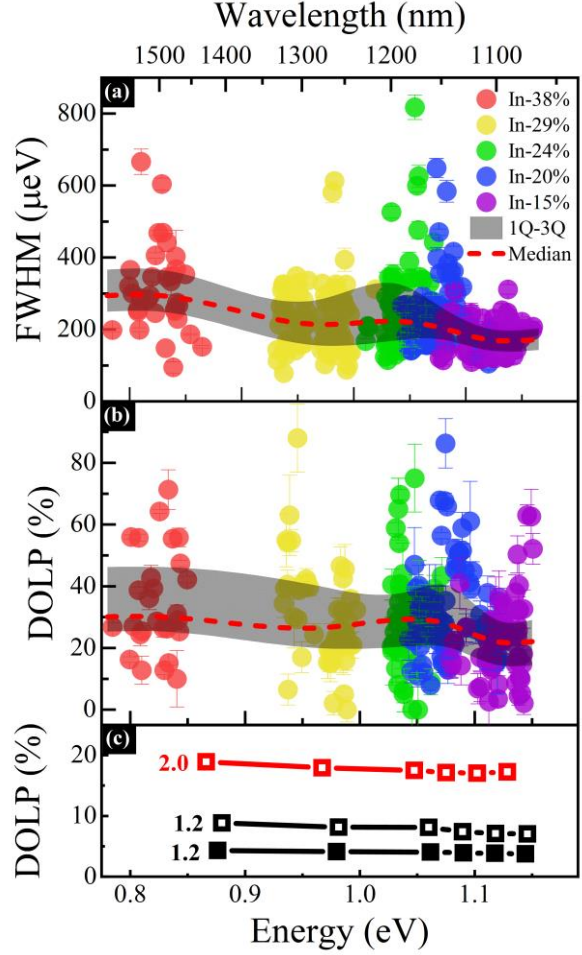

FIG. 3. (a) The linewidth of single QD emission lines as a function of emission energy for investigated structures from 15% to 38% MBL indium content (marked range between the first and third quartile and red dashed line median for individual samples). (b) DOLP values for single QD emission lines in a function of emission energy for the investigated structures. (c) Calculated DOLP values as a function of MBL indium content for various QD parameters: H = 5 nm, D = 40 nm (solid squares), H = 6 nm, D = 30 nm (open squares) with different lateral aspect ratios.

Polarization of the QD emission for the light propagating perpendicularly to the sample surface is mainly defined by the QD in-plane asymmetry through the effects of light and heavy hole states mixing and electron-hole exchange interaction. The inequality of oscillator strengths for the orthogonal in-plane axes results in some degree of linear polarization: $\text{DOLP} = (I_{\max} - I_{\min})/(I_{\max} + I_{\min})$ [79]. For QDs with significantly decreased strain (like those studied here), asymmetric growth along the crystallographic directions [110] and [1-10] is often observed due to the enhanced sensitivity to the atomic steps and hence more anisotropic diffusion during the growth of QDs, resulting in the increased DOLP values [79–82]. For our QDs on MBL, we observe similar median DOLP values of 20-30% for the entire spectral range with a significant spread of values (from near 0% to almost 90%). It is most likely due to some degree of randomness in the orientation of elongation axes in the QD ensemble [79,83]. Figure 3(b) shows the measured DOLP values for all structures (using standard fitting procedure [84]). The observed values of DOLP close to 100% can be related to the QDs with high contribution of the light hole states. Increasing the indium content in the MBL layer reduces the strain between the base material and the QD layer, which can result in the formation of QDs with increased asymmetry. However, the lack of significant DOLP enhancement for longer wavelengths and our subsequent results for excitonic complexes suggest no substantial changes in the asymmetry of the studied QDs. Based on our calculations, even for QDs with only slight asymmetry (LAR = 1.2), a significant mixing of hole states is expected. It is partly due to reduced strain and hence decreased splitting of the hole bands. Additionally, particularly for the hole, the reduction of indium concentration contrast between the top of MBL and a QD makes the binding potential significantly shallower leading to stronger wavefunction leakage into the interface where the heavy- and light-hole bands cross. This enhancement of hole subband mixing results in DOLP above 7% for a QD with $H = 6$ nm and $D = 30$ nm (see Fig. 3(c)). Calculated DOLP is lower (4%) for a flatter QD with $H = 5$ nm and $D = 40$ nm. This result suggests that the former geometry is closer to the actual dimensions of the investigated QDs. Let us recall that there were no significant differences in the emission energy between both QD geometries (see Fig. 2(c)). Enhancing the QD asymmetry can further increase the DOLP value up to 18% for structures with LAR = 2.0, as shown in Fig. 3(c). Additionally, the strain in a QD can be overestimated in theory, as it can be additionally relaxed in an actual structure due to some defects located at the top of MBL. This could explain the underestimation of DOLP in our modeling. Overall, the change in the QD emission energy, resulting from modifying the MBL indium content, is not accompanied by a significant change in DOLP. We attribute this result to the change in the indium concentration in the dots on In-rich MBLs partly restoring the indium composition contrast between the top of MBL and a QD, which tends to decrease for In-rich structures.

### B. Quantum dot energy level structure by photoluminescence excitation

Next, we investigate the QD energy level structure using photoluminescence excitation spectroscopy. The black line in Fig. 4(a) shows an exemplary emission spectrum for a QD emitting in the third telecom window (In-38% structure). The PLE spectrum for the emission marked with the red arrow is plotted in red. Positions of PLE maxima provide insight into the structure of QD excited states. We also observe an agreement between PL and PLE spectra (green arrow in Fig. 4(a)), which usually indicates the same excited state seen in both the emission and absorption-like experiments. The knowledge on bright excited QD states allows for quasi-resonant excitation. By pumping into an excited state, we avoid generating carriers in the barrier and MBL, significantly decreasing the probability of feeding charge trap states. It should decrease the PL linewidth for quasi-resonant excitation compared to the non-resonantly excited PL. In Fig. 4(a), we observe such a reduction of linewidth from 640 μeV for non-resonant excitation to 460 μeV for quasi-resonant excitation (blue line) into one of the QD excited states with the energy of 0.838 eV marked with the blue arrow in Fig. 4(a).

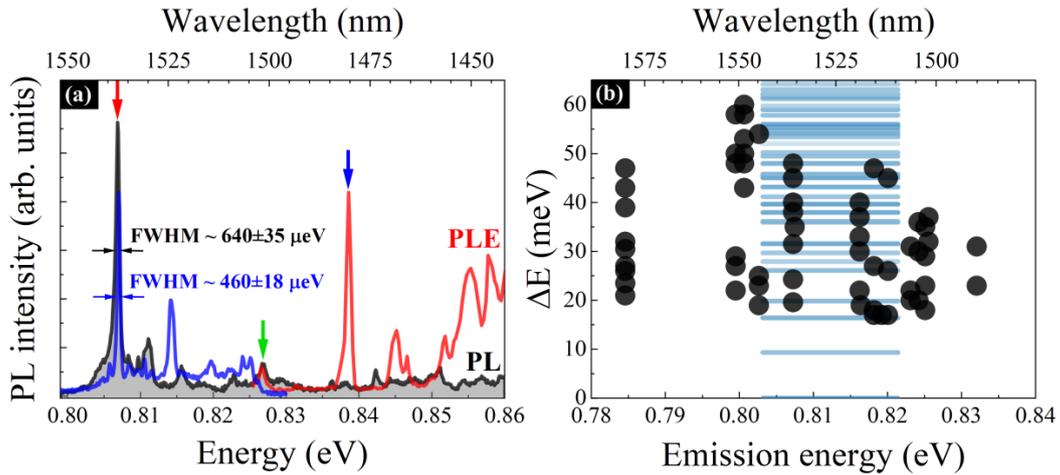

FIG. 4. (a) PL spectrum from a single QD for the In-38% structure for non-resonant excitation (black line). PLE spectrum (red line) for the indicated emission line (red arrow). PL spectra (blue line) based on the quasi-resonant excitation at the indicated PLE resonance (blue arrow). Observation of a signal for higher QD state in both PL and PLE spectra (green arrow). (b) The relative energy of excited QD states from single-dot PLE spectra with the underlaid simulated levels (blue lines with intensities proportional to the oscillator strength).

The observed linewidth decrease is slight, and still, the emission line is relatively broad. It shows that the charge traps expected in the MBL, bypassed, at least partly, during the quasi-resonant excitation, do not dominate the observed QD linewidths. Lower than expected reduction of PL linewidth may result from built-in excess carriers in the structure (unintentional doping). Another reason may be the formation of charge traps at energies close to the QD emitting states, which are also fed during the quasi-resonant excitation. For QDs emitting in the third telecom window, we obtain a series of absorption maxima (energies given relative to the QD emission energy). We show them in Fig. 4(b), where we observe a dense structure of higher-energy states with the underlaid simulated levels (blue lines with intensities proportional to the oscillator strength). The measurement range for the excitation-emission energy difference is experimentally limited due to the inability to suppress the excitation laser line at small energy differences (below $\Delta E \approx 15$ meV) to the emission line. Both experiments and simulations show a similarly dense structure of bright excited states. A dense ladder of excited states is expected in QDs grown in lowered-strain conditions provided by the In-rich MBL, making the confinement weaker and the dots relatively large. It also agrees with the observed limited narrowing of the QD emission line for quasi-resonant excitation. In the investigated dots, we still excite much higher excited states than in typical InAs/GaAs QDs where the energy ladder is sparser.

### C. Excitonic complexes

As the QD emission is directly defined by the excitonic complexes it confines, we look in more detail at their properties. We characterize exciton and biexciton complexes in all investigated structures with high-spatial-resolution PL measurements. Figure 5(a) shows exemplary spectra of excitation-power dependence of exciton (X) and biexciton (XX) radiative recombination in a single QD emitting in the third telecom window (structure In-38%). We identify the emission lines based on their power-dependent intensities, linear for the exciton and almost quadratic for the biexciton (Fig. 5(b)) [85]. Additionally, we confirm their origin in the anti-phase dependence of the energy position as a function of the linear polarization angle (Fig. 5(c)) [86].

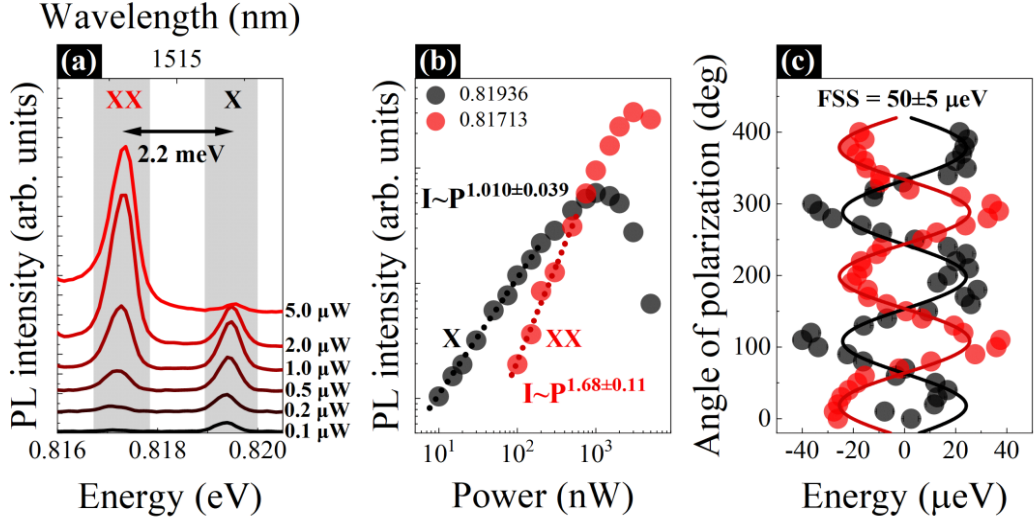

FIG. 5. (a) PL spectra from a single QD for non-resonant excitation (sample In-38%) for the excitation power values from 0.1 to 5 μW. (b) The intensity of the exciton (X) and biexciton (XX) lines in a function of excitation power with power-function fits (lines). (c) Emission energy of X and XX lines for different linear polarization angles with sine fits (lines).

By fitting to the low-excitation-power part of dependences, we find the exponents of 1.01±0.04 for the exciton and 1.68±0.11 for the biexciton. We find no clear dependence between the exponents and emission energy based on all investigated X-XX pairs. The average exponent for X is 0.948±0.031 and 1.755±0.050 for XX. Their ratio is about 1.85, i.e. it is not reaching the typical value of 2 expected for QDs in the strong confinement regime [85], suggesting intermediate confinement regime, which is expected given the dense ladder of hole states with levels' separations lower than the electron-hole Coulomb interaction of ~15 meV.

Biexciton binding energy, determined from the energy separation between the X and XX lines, for the discussed pair is about 2.2 meV. Figure 6(a) shows the values of the biexciton binding energy for all examined X-XX pairs. We observe a slight change with the emission energy (reduction of the binding energy for longer wavelengths). The range of values within a QD ensemble is relatively broad (over 1.5 meV in the most dispersed case), whereas the changes between different structures are less pronounced. It confirms the very weak dependence of XX binding energy on the MBL composition. The obtained XX binding energies ranging from 1.06 to 2.94 meV are comparable to those for standard GaAs-based QDs [87,88] and those grown on MBL by other groups [37,89]. We compare the experimental values with simulations for QDs ($H$ = 6 nm, $D$ = 30 nm) with different LAR. We show this in Fig. 6(a), where good agreement may be seen.

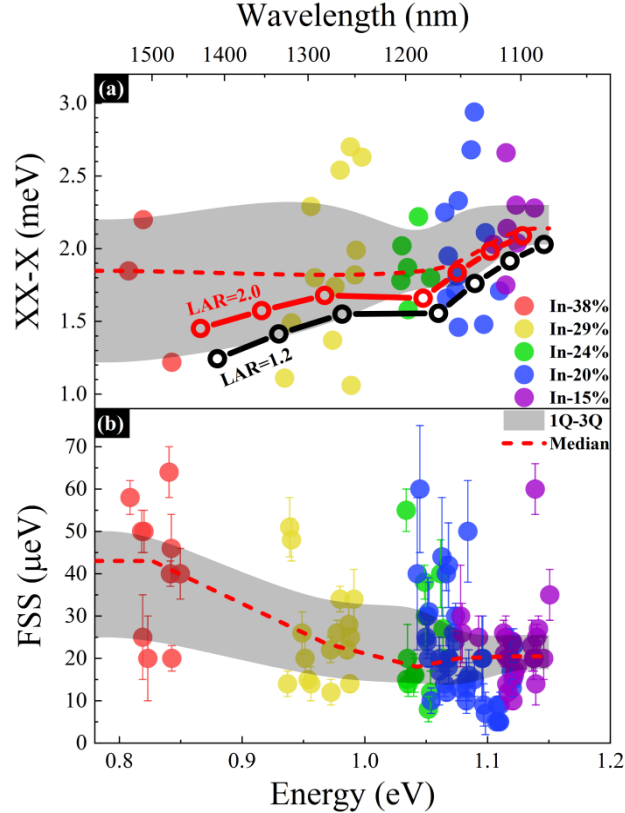

FIG. 6. (a) XX binding energies (solid points) as a function of emission energy with a marked range between the first and third quartile and dashed red line median for individual samples. Theoretical simulations of binding energies for QDs ($H$ = 6 nm, $D$ = 30 nm) with different LAR (open points). (b) The fine structure splitting of single QD emission lines as a function of emission energy (marked range between the first and third quartile and dashed red line median for individual samples).

Based on the polarization dependence of X and XX energies (Fig. 5(c)), we determine the fine structure splitting (FSS) of 50±5 µeV for the selected case, suggesting at least some QD confinement potential asymmetry. All the collected FSS values are presented in Fig. 6(b). They range from 5 to 64 µeV, with some tendency of higher FSS for longer wavelengths, with however their accuracy limited by relatively large inhomogeneous linewidths. The observed weak trend may indicate an increase in the asymmetry of the dots grown in lower strain conditions (higher indium content in MBL). Such increased asymmetry could, in turn, be connected to high DOLP observed for many lines with a slight increase with the redshift of emission. However, in our case, the FSS trend is barely noticeable and the obtained values are relatively low in the entire spectral range (especially for the wavelengths below 1350 nm), which usually corresponds to rather in-plane symmetric nanostructures, or symmetric confinement potential for at least one of the carriers. In addition, we do not include in Fig. 6(b) cases with almost zero FSS. These would be difficult to distinguish from the charged exciton emission lines in polarization-resolved analysis, considering the finite spectral resolution of our setup of about 20 µeV, and the limitations of the analysis of the results (about 5 μeV in fitting procedure and lines' separation). Therefore, the actual FSS values (especially for shorter wavelengths) can be even lower.

**D. Carrier dynamics**

The PL decay time of QD emission is primarily related to the radiative lifetime of the confined carriers, especially at sufficiently low temperatures, when the contribution from the non-radiative processes is minimized. The radiative lifetime directly corresponds to the oscillator strength of optical transitions, related to the confinement potential in a QD. It is crucial for implementing single-photon sources based on QDs as it impacts the maximal frequency of operation of such devices. Our QDs are characterized by either fast mono-exponential decays or a combination of short and long decay times, as shown in Fig. 7(a) and 7(b), respectively, with exemplary results showing the fast and slow decay components.

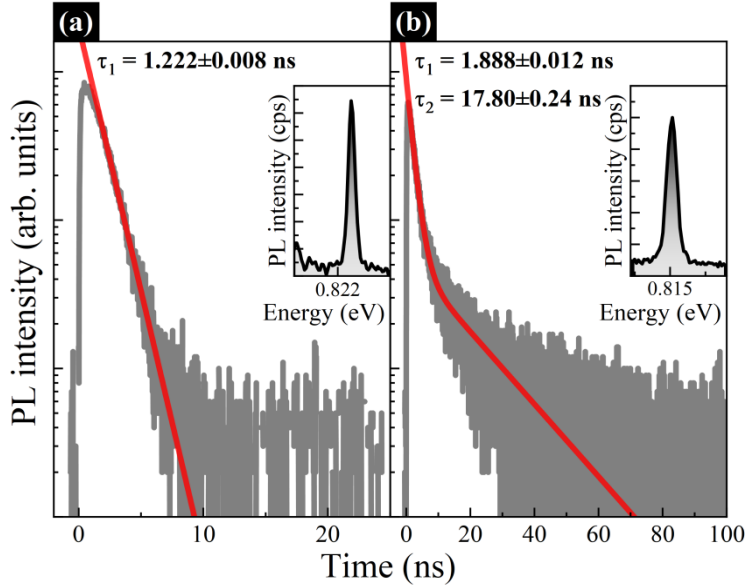

FIG. 7. TRPL results for (a) fast single exponential decay and (b) combination of short and long decay times with additional exponential fit (red line) for single QD (structure In-38%) non-resonantly excited.

The mean values of the measured decay times (non-resonant pulsed excitation) for many investigated lines (different complexes; uncategorized) and identified exciton and biexciton complexes are presented in Table II. In most cases, two decay times are observed, especially for QDs emitting above 1200 nm, which may indicate the presence of other processes manifested by slow decay component for structures with increased indium content in the top part of the MBL layer. They may be related to the influence of the local QD charge environment, i.e additional energy states likely associated with neigbouring charge-trapping defects from which a delayed refilling of QDs can occur. The concentration of such charge traps in the MBL can be elevated for higher indium content. It agrees with the observed linewidth increase for QDs emitting at longer wavelengths, as discussed above. Increased influence of defect states can also result in the observation of the additional decay time. Two decay times have been seen for analogous QD structures emitting in the third telecom window and fabricated by MOVPE confirming a similar impact of the charge traps in MBL [51].

TABLE II. Decay times of single QD emission lines (mean values) for all investigated structures divided into the spectral ranges of the emission. $\tau_1$ and $\tau_2$ cover many QD lines of different complexes, while $\tau_X$ and $\tau_{XX}$ are for the short decay time for identified X and XX lines only.

|  | < 1200 nm | 1200 – 1400 nm | 1400 – 1600 nm |
| --- | --- | --- | --- |
| Short decay time (ns) $\tau_1$ | 1.15±0.16 | 1.432±0.081 | 1.52±0.48 |
| Long decay time (ns) $\tau_2$ | not observed | 5.20±0.68 | 6.4±1.3 |
| X decay time (ns) $\tau_X$ | 1.56±0.35 | 2.26±0.40 | 2.64±0.43 |
| XX decay time (ns) $\tau_{XX}$ | 0.81±0.15 | 1.10±0.20 | 1.39±0.12 |
| Ratio $\tau_X/\tau_{XX}$ | 1.86-2.00 | 1.81-2.45 | 1.73-2.04 |

We observe a slight increase in the short decay times (for the uncategorized QD lines of different complexes) for longer wavelengths, from an average of 1.15 ns for QDs emitting below 1200 nm to 1.52 ns for QDs emitting in the third telecom window (1450-1600 nm). This change suggests lower oscillator strength for In-rich MBL dots. Comparable trends were observed for the identified X and XX complexes (see Table II). Information on their lifetimes is relevant due to excluding other complexes from the decay time statistics. We find an average exciton lifetime of 1.56 ns for QDs emitting below 1200 nm and 2.64 ns for dots in the third telecom window. These times are slightly longer than typically observed in InAs QDs emitting below 1000 nm [9] and for similar structures grown by MOVPE [41]. Similarly, we observe increased X and XX decay times for longer wavelengths. It is worth noting that for the identified X and XX emission, a slow decay component is also present for most of the lines. Table II also presents the exciton/biexciton lifetime ratio, often used to identify the QD exciton confinement regime. We observe $\tau_X/\tau_{XX}$ ratios around 1.8, i.e., below 2 for all the spectral ranges and all MBL compositions, suggesting intermediate confinement. This agrees with dense confined states ladder, especially for holes. For QDs emitting in the second (third) telecom window, the separation is ~21 meV (~15 meV) for electron states and ~4.3 meV (~3.8 meV) for hole states (for a QD with H = 6 nm and D = 30 nm, LAR = 1.2). The separation of single-particle states compared to the exciton binding energy of about 12-15 meV confirms a weaker confinement potential and is characteristic for the intermediate confinement regime.

### E. Single-photon emission in the telecom range

From the photon auto-correlation measurements for the exciton emission, we obtain the second-order correlation functions $g^{(2)}(\tau)$ under non-resonant cw excitation (640 nm), which give an insight into the quality of single photon generation. Figures 8(a) and 8(b) show exemplary results in the second and third telecom windows. For most of the investigated cases, the as-measured values of $g^{(2)}(0)$ are below 0.2, demonstrating the telecom-range single-photon emission from our QDs with good source purity, regardless of the spectral range and modifications in the MBL – see Fig. 8(c). Fitting the experimental data [46] provides even lower values, marked by rimmed symbols in Fig. 8(c). The determined $g^{(2)}(0)$ values are comparable with those for InAs/InP and InAs/InGaAlAs/InP QDs (both symmetric and asymmetric) obtained under non-resonant excitation [15,21,28,90–92], however still worse than the best reported for other GaAs-based dots [12,20]. The level of signal detected from the investigated QDs is still not very high (below 5000 counts/s for the best cases at saturation using single-photon detectors), which results in limited accuracy of $g^{(2)}(0)$ determination, which can be a limitation in precise derivation of very low $g^{(2)}(0)$ values. It can be improved when the material structural quality is improved and hence contribution of non-radiative losses is decreased.

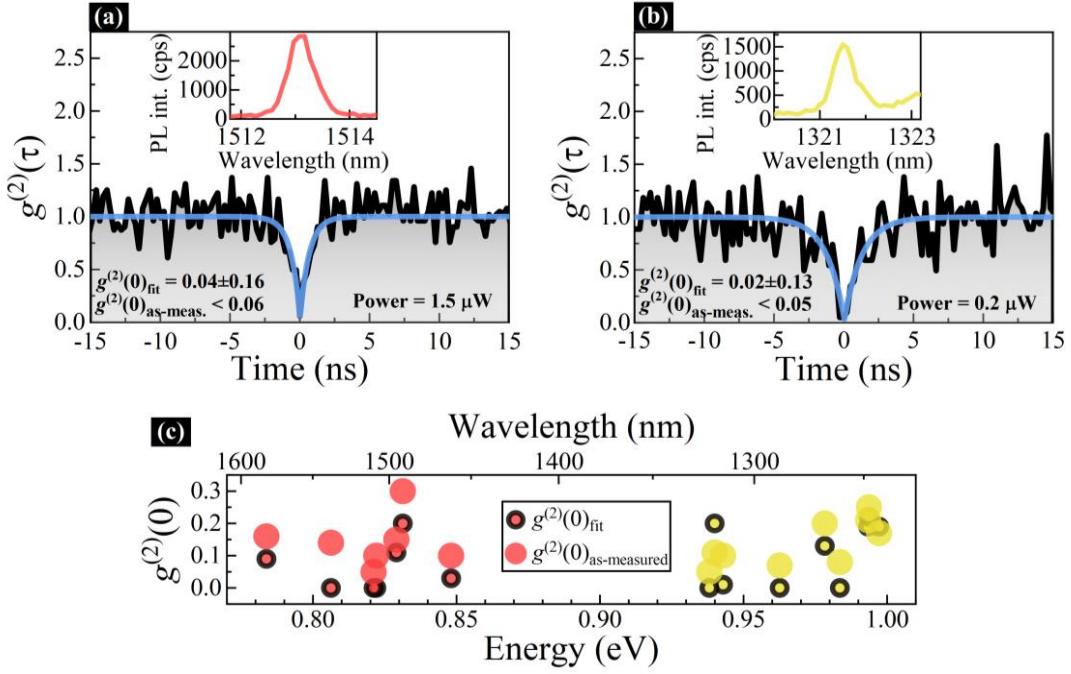

FIG. 8. Second-order correlation function $g^{(2)}(\tau)$ for X emission under cw excitation with 640 nm line (a) for structure In-38% and (b) for structure In-29%. The blue line is the fit to the experimental data, and the insets show the emission spectra of the investigated lines. (c) Values for all investigated cases from both structures.

The $g^{(2)}$ measurements under pulsed excitation allow us to demonstrate single-photon triggered QD emission in both the second and third telecom windows, as shown in Fig. 9(a) and 9(b). The limitation of these experiments was a relatively weak signal in these conditions, which forced us to use higher excitation when compared to cw measurements, plus the existence of the long decay time component in PL. As a consequence of the latter, the counts in between the pulses do not reach the zero level (or the detector dark counts level), which is even more severe for the 1.5 μm (In-38%) case. Therefore, we used a twice smaller excitation repetition then (see Fig. 9(a)). The near saturation excitation conditions do not allow us to directly connect the fitted decay times to the result of dynamics characterization obtained for much lower average powers. Independently of these, we observe a significant signal reduction for zero time delay indicating a clear single-photon emission in a triggered mode. The nonideal suppression of multiphoton events for small delays (see corresponding zoom-in for $g^{(2)}(\tau)$ in Fig. 9(c) and 9(d)) is additionally hindered by observation of the typical fingerprint of the carrier recapture processes from the charge states surrounding the dots [16,93,94]. To account for the long decay time component, two decay times were included in the fitting formula [95], which was, however, insufficient to reconstruct properly the near zero delay range, where inclusion of the recapturing was necessary (after Ref. [16]). Based on the fitting with this dependence, we obtain $g^{(2)}(0)$ values of 0.07 and 0.17 for the 1.3 and 1.5 μm cases, respectively, and the natural $g^{(2)}(\tau)$ background on a similar level of about 0.05-0.1, estimated to be the detector dark counts. Despite the mentioned limitations, this single photon emission purity indicates the application potential of these QDs, where the improvement of the characteristics should be possible with resonant excitation and further optimizations in the growth procedure. Optimization of MBL growth parameters, as well as the capping layer, should allow reducing the number of defects in the structure including the vicinity of QDs [45,78] and hence suppressing the observed recapturing processes and the occurrence of slow decay component in the PL dyanamics.

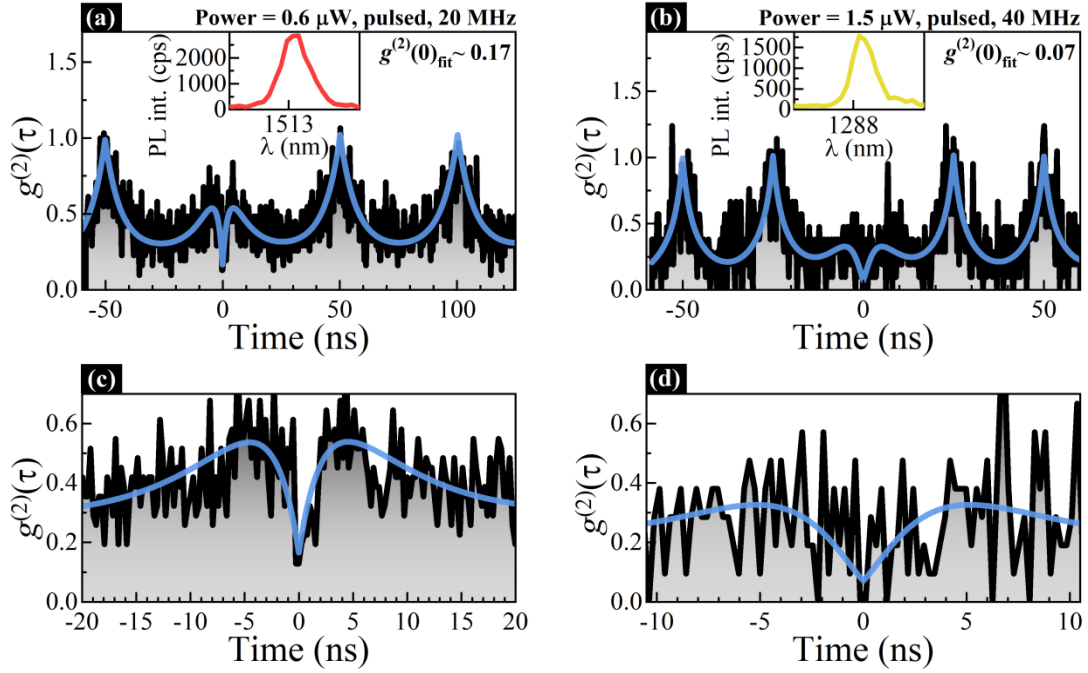

FIG. 9. Second-order correlation function $g^{(2)}(\tau)$ for X emission under pulsed excitation with 805 nm line (a) for structure In-38% and (b) for structure In-29%. The blue line is the fit to the experimental data based on the additional recapture process and two-component decay, and the insets show the emission spectra of the investigated lines. Zoom-in ranges for small delays for $g^{(2)}(\tau)$ dependence and fit (c) for the third telecom range and (d) for the second telecom range, respectively.

## IV. CONCLUSION

We presented a study of single InGaAs QDs that were grown by MBE on InGaAs metamorphic buffers, on GaAs substrate. By changing the indium concentration in the top part of MBL, we redshifted the QD emission and reached the third telecom window. We showed that the emission energy changes due to the reduced strain in QDs but also due to the varying indium content within the dots, induced by the composition changes in the MBL. We identified linewidths for many single QDs showing a slight increase for QDs on In-rich MBLs, suggesting a rise in the density of charge traps near QDs. In the polarization-resolved experiments, we obtained no fingerprint of the significant asymmetry change for different MBL compositions. The QD excited states could directly be probed in single dots by high-spatial-resolution photoluminescence excitation spectroscopy, allowing for the demonstration of the QD emission linewidth decrease for quasi-resonant excitation. We identified basic excitonic complexes and followed the changes in biexciton binding energy and fine structure splitting for different MBL compositions, suggesting elongation of the dots emitting at longer wavelengths. The single-dot radiative lifetimes suggested a slight QD oscillator strength decrease for those on In-rich MBLs, whereas the ratios of the exciton to biexciton lifetimes and the slopes in the low excitation power dependences indicate an intermediate confinement regime in these dots. We also demonstrated single photon emission in the second and third telecom windows with the $g^{(2)}(0)$ values comparable with InP-based QDs in the same range. This material system, grown by MBE, is relatively new compared to the more established MOVPE-grown solutions, but it already shows its competitiveness as a platform for telecom applications, serving also as a good alternative to the InP-based QDs. Our results showed routes for further developments and optimizations, which should primarily concern improvements in the crystalographic quality of the MBL and reduction of the defects concentration, which will affect many of the crucial properties, as e.g. minimizing the contribution of slow PL decay component or decreasing the inhomogenous linewidth which in turn will improve the accuracy of FSS and $g^{(2)}(0)$ determination.

## ACKNOWLEDGMENTS

We would like to thank Monika Emmerling (University of Würzburg) for the fabrication of mesas and apertures on the sample surface. We also thank Anna Musiał (Wrocław University of Science and Technology) for assistance with the optical spectroscopy setups. We are grateful to Krzysztof Gawarecki (Wrocław University of Science and Technology) for sharing his implementation of the $\mathbf{k}\cdot\mathbf{p}$ method. Numerical calculations have been carried out using resources provided by Wroclaw Centre for Networking and Supercomputing (https://wcss.pl).

## FUNDING

P.A.W., F.J. and S.H. acknowledge financial support by the free state of Bavaria. S.H. and P.A.W. acknowledge financial support by the H2020 Marie Skłodowska-Curie Actions under ITN project 4PHOTON (Grant Agreement ID: 721394), P.W. acknowledges financial support by the European Union under the European Social Fund.

## AUTHOR CONTRIBUTIONS

Conceptualization: P.P. and G.S.; methodology: P.W., M.G, P.P., M.P. and S.G.; investigation: P.W., M.G., M.P. and S.G.; formal analysis: P.W., M.G., P.P. and G.S.; funding acquisition: G.S., S.H.; project administration: P.P. and G.S.; software: M.G.; writing – original draft preparation: P.W.; writing – review and editing: all authors; resources: P.A.W., F.J. and S.H; supervision: P.P. and G.S.

## CONFLICT OF INTEREST

The authors declare no conflict of interest.

[1] C. Shang, Y. Wan, J. Selvidge, E. Hughes, R. Herrick, K. Mukherjee, J. Duan, F. Grillot, W. W. Chow, and J. E. Bowers, *Perspectives on Advances in Quantum Dot Lasers and Integration with Si Photonic Integrated Circuits*, ACS Photonics **8**, 2555 (2021).

[2] Z. Yao, C. Jiang, X. Wang, H. Chen, H. Wang, L. Qin, and Z. Zhang, *Recent Developments of Quantum Dot Materials for High Speed and Ultrafast Lasers*, Nanomaterials **12**, 7 (2022).

[3] A. Yadav, N. B. Chichkov, E. A. Avrutin, A. Gorodetsky, and E. U. Rafailov, *Edge Emitting Mode-Locked Quantum Dot Lasers*, Progress in Quantum Electronics **87**, 100451 (2023).

[4] S. Bauer, V. Sichkovskyi, O. Eyal, T. Septon, A. Becker, I. Khanonkin, G. Eisenstein, and J. P. Reithmaier, *1.5-µm Indium Phosphide-Based Quantum Dot Lasers and Optical Amplifiers: The Impact of Atom-Like Optical Gain Material for Optoelectronics Devices*, IEEE Nanotechnology Magazine **15**, 23 (2021).

[5] M. Z. M. Khan, T. K. Ng, and B. S. Ooi, *Self-Assembled InAs/InP Quantum Dots and Quantum Dashes: Material Structures and Devices*, Progress in Quantum Electronics **38**, 237 (2014).

[6] N. Ozaki, K. Takeuchi, S. Ohkouchi, N. Ikeda, Y. Sugimoto, H. Oda, K. Asakawa, and R. A. Hogg, *Monolithically Grown Multi-Color InAs Quantum Dots as a Spectral-Shape-Controllable near-Infrared Broadband Light Source*, Applied Physics Letters **103**, 051121 (2013).

[7] L. Seravalli, M. Gioannini, F. Cappelluti, F. Sacconi, G. Trevisi, and P. Frigeri, *Broadband Light Sources Based on InAs/InGaAs Metamorphic Quantum Dots*, Journal of Applied Physics **119**, 143102 (2016).

[8] P. Holewa, M. Gawełczyk, A. Maryński, P. Wyborski, J. P. Reithmaier, G. Sęk, M. Benyoucef, and M. Syperek, *Optical and Electronic Properties of Symmetric InAs/(In,Al,Ga)As/InP Quantum Dots Formed by Ripening in Molecular Beam Epitaxy: A Potential System for Broad-Range Single-Photon Telecom Emitters*, Phys. Rev. Appl. **14**, 064054 (2020).

[9] S. Buckley, K. Rivoire, and J. Vučković, *Engineered Quantum Dot Single-Photon Sources*, Rep. Prog. Phys. **75**, 126503 (2012).

[10] P. Senellart, G. Solomon, and A. White, *High-Performance Semiconductor Quantum-Dot Single-Photon Sources*, Nature Nanotech **12**, 11 (2017).

[11] X. Cao, M. Zopf, and F. Ding, *Telecom Wavelength Single Photon Sources*, J. Semicond. **40**, 071901 (2019).

[12] Y. Arakawa and M. J. Holmes, *Progress in Quantum-Dot Single Photon Sources for Quantum Information Technologies: A Broad Spectrum Overview*, Applied Physics Reviews **7**, 021309 (2020).

[13] C. Schimpf, M. Reindl, F. Basso Basset, K. D. Jöns, R. Trotta, and A. Rastelli, *Quantum Dots as Potential Sources of Strongly Entangled Photons: Perspectives and Challenges for Applications in Quantum Networks*, Applied Physics Letters **118**, 100502 (2021).

[14] S. Liu, K. Srinivasan, and J. Liu, *Nanoscale Positioning Approaches for Integrating Single Solid-State Quantum Emitters with Photonic Nanostructures*, Laser & Photonics Reviews **15**, 2100223 (2021).

[15] A. Musiał, M. Mikulicz, P. Mrowiński, A. Zielińska, P. Sitarek, P. Wyborski, M. Kuniej, J. P. Reithmaier, G. Sęk, and M. Benyoucef, *InP-Based Single-Photon Sources Operating at Telecom C-Band with Increased Extraction Efficiency*, Applied Physics Letters **118**, 221101 (2021).

[16] P. Holewa, A. Sakanas, U. M. Gür, P. Mrowiński, A. Huck, B.-Y. Wang, A. Musiał, K. Yvind, N. Gregersen, M. Syperek, and E. Semenova, *Bright Quantum Dot Single-Photon Emitters at Telecom Bands Heterogeneously Integrated on Si*, ACS Photonics **9**, 2273 (2022).

[17] F. Olbrich, J. Kettler, M. Bayerbach, M. Paul, J. Höschele, S. L. Portalupi, M. Jetter, and P. Michler, *Temperature-Dependent Properties of Single Long-Wavelength InGaAs Quantum Dots Embedded in a Strain Reducing Layer*, Journal of Applied Physics **121**, 184302 (2017).

[18] Ł. Dusanowski, M. Syperek, J. Misiewicz, A. Somers, S. Höfling, M. Kamp, J. P. Reithmaier, and G. Sęk, *Single-Photon Emission of InAs/InP Quantum Dashes at 1.55 μm and Temperatures up to 80 K*, Applied Physics Letters **108**, 163108 (2016).

[19] A. W. Elshaari, W. Pernice, K. Srinivasan, O. Benson, and V. Zwiller, *Hybrid Integrated Quantum Photonic Circuits*, Nat. Photonics **14**, 5 (2020).

[20] S. L. Portalupi, M. Jetter, and P. Michler, *InAs Quantum Dots Grown on Metamorphic Buffers as Non-Classical Light Sources at Telecom C-Band: A Review*, Semicond. Sci. Technol. **34**, 053001 (2019).

[21] P. Holewa, S. Kadkhodazadeh, M. Gawełczyk, P. Baluta, A. Musiał, V. G. Dubrovskii, M. Syperek, and E. Semenova, *Droplet Epitaxy Symmetric InAs/InP Quantum Dots for Quantum Emission in the Third Telecom Window: Morphology, Optical and Electronic Properties*, Nanophotonics **11**, 1515 (2022).

[22] A. Barbiero, J. Huwer, J. Skiba-Szymanska, D. J. P. Ellis, R. M. Stevenson, T. Müller, G. Shooter, L. E. Goff, D. A. Ritchie, and A. J. Shields, *High-Performance Single-Photon Sources at Telecom Wavelength Based on Broadband Hybrid Circular Bragg Gratings*, ACS Photonics **9**, 3060 (2022).

[23] M. D. Birowosuto, H. Sumikura, S. Matsuo, H. Taniyama, P. J. van Veldhoven, R. Nötzel, and M. Notomi, *Fast Purcell-Enhanced Single Photon Source in 1,550-μm Telecom Band from a Resonant Quantum Dot-Cavity Coupling*, Sci Rep **2**, 1 (2012).

[24] X. Liu, K. Akahane, N. A. Jahan, N. Kobayashi, M. Sasaki, H. Kumano, and I. Suemune, *Single-Photon Emission in Telecommunication Band from an InAs Quantum Dot Grown on InP with Molecular-Beam Epitaxy*, Applied Physics Letters **103**, 061114 (2013).

[25] Ł. Dusanowski, M. Syperek, P. Mrowiński, W. Rudno-Rudziński, J. Misiewicz, A. Somers, S. Höfling, M. Kamp, J. P. Reithmaier, and G. Sęk, *Single Photon Emission at 1.55 μm from Charged and Neutral Exciton Confined in a Single Quantum Dash*, Appl. Phys. Lett. **105**, 021909 (2014).

[26] K. Takemoto, Y. Nambu, T. Miyazawa, Y. Sakuma, T. Yamamoto, S. Yorozu, and Y. Arakawa, *Quantum Key Distribution over 120 Km Using Ultrahigh Purity Single-Photon Source and Superconducting Single-Photon Detectors*, Sci Rep **5**, 1 (2015).

[27] T. Müller, J. Skiba-Szymanska, A. B. Krysa, J. Huwer, M. Felle, M. Anderson, R. M. Stevenson, J. Heffernan, D. A. Ritchie, and A. J. Shields, *A Quantum Light-Emitting Diode for the Standard Telecom Window around 1,550 μm*, Nat Commun **9**, 1 (2018).

[28] P. Wyborski, A. Musiał, P. Mrowiński, P. Podemski, V. Baumann, P. Wroński, F. Jabeen, S. Höfling, and G. Sęk, *InP-Substrate-Based Quantum Dashes on a DBR as Single-Photon Emitters at the Third Telecommunication Window*, Materials **14**, 4 (2021).

[29] M. Anderson, T. Müller, J. Huwer, J. Skiba-Szymanska, A. B. Krysa, R. M. Stevenson, J. Heffernan, D. A. Ritchie, and A. J. Shields, *Quantum Teleportation Using Highly Coherent Emission from Telecom C-Band Quantum Dots*, Npj Quantum Inf **6**, 1 (2020).

[30] M. Anderson, T. Müller, J. Skiba-Szymanska, A. B. Krysa, J. Huwer, R. M. Stevenson, J. Heffernan, D. A. Ritchie, and A. J. Shields, *Coherence in Single Photon Emission from Droplet Epitaxy and Stranski–Krastanov Quantum Dots in the Telecom C-Band*, Applied Physics Letters **118**, 014003 (2021).

[31] L. Wells, T. Müller, R. M. Stevenson, J. Skiba-Szymanska, D. A. Ritchie, and A. J. Shields, *Coherent Light Scattering from a Telecom C-Band Quantum Dot*, arXiv:2205.07997.

[32] G. Shooter, Z.-H. Xiang, J. R. A. Müller, J. Skiba-Szymanska, J. Huwer, J. Griffiths, T. Mitchell, M. Anderson, T. Müller, A. B. Krysa et al., *1 GHz Clocked Distribution of Electrically Generated Entangled Photon Pairs*, Opt. Express, OE **28**, 36838 (2020).

[33] M. Anderson, T. Müller, J. Huwer, J. Skiba-Szymanska, A. B. Krysa, R. M. Stevenson, J. Heffernan, D. A. Ritchie, and A. J. Shields, *Gigahertz-Clocked Teleportation of Time-Bin Qubits with a Quantum Dot in the Telecommunication C Band*, Phys. Rev. Appl. **13**, 054052 (2020).

[34] L. Seravalli, M. Minelli, P. Frigeri, P. Allegri, V. Avanzini, and S. Franchi, *The Effect of Strain on Tuning of Light Emission Energy of InAs/InGaAs Quantum-Dot Nanostructures*, Applied Physics Letters **82**, 2341 (2003).

[35] A. R. Kovsh, A. E. Zhukov, N. A. Maleev, S. S. Mikhrin, V. M. Ustinov, A. F. Tsatsul'nikov, M. V. Maksimov, B. V. Volovik, D. A. Bedarev, Yu. M. Shernyakov et al., *Lasing at a Wavelength Close to 1.3 µm in InAs Quantum-Dot Structures*, Semiconductors **33**, 929 (1999).

[36] R. Heitz, I. Mukhametzhanov, A. Madhukar, A. Hoffmann, and D. Bimberg, *Temperature Dependent Optical Properties of Self-Organized InAs/GaAs Quantum Dots*, J. Electron. Mater. **28**, 520 (1999).

[37] C. Carmesin, F. Olbrich, T. Mehrtens, M. Florian, S. Michael, S. Schreier, C. Nawrath, M. Paul, J. Höschele, B. Gerken et al., *Structural and Optical Properties of InAs/(In)GaAs/GaAs Quantum Dots with Single-Photon Emission in the Telecom C-Band up to 77 K*, Phys. Rev. B **98**, 125407 (2018).

[38] L. H. Li, M. Rossetti, G. Patriarche, and A. Fiore, *Growth of InAs Bilayer Quantum Dots for Long-Wavelength Laser Emission on GaAs*, Journal of Crystal Growth **301–302**, 959 (2007).

[39] B. Alloing, C. Zinoni, L. H. Li, A. Fiore, and G. Patriarche, *Structural and Optical Properties of Low-Density and In-Rich InAs/GaAs Quantum Dots*, Journal of Applied Physics **101**, 024918 (2007).

[40] P. Wyborski, P. Podemski, P. A. Wroński, F. Jabeen, S. Höfling, and G. Sęk, *Electronic and Optical Properties of InAs QDs Grown by MBE on InGaAs Metamorphic Buffer*, Materials **15**, 3 (2022).

[41] M. Paul, F. Olbrich, J. Höschele, S. Schreier, J. Kettler, S. L. Portalupi, M. Jetter, and P. Michler, *Single-Photon Emission at 1.55 Mm from MOVPE-Grown InAs Quantum Dots on InGaAs/GaAs Metamorphic Buffers*, Applied Physics Letters **111**, 033102 (2017).

[42] E. S. Semenova, A. E. Zhukov, S. S. Mikhrin, A. Y. Egorov, V. A. Odnoblyudov, A. P. Vasil'ev, E. V. Nikitina, A.R. Kovsh, N. V. Kryzhanovskaya, A. G. Gladyshev et al., *Metamorphic Growth for Application in Long-Wavelength (1.3–1.55 Mm) Lasers and MODFET-Type Structures on GaAs Substrates*, Nanotechnology **15**, S283 (2004).

[43] E. S. Semenova, R. Hostein, G. Patriarche, O. Mauguin, L. Largeau, I. Robert-Philip, A. Beveratos, and A. Lemaître, *Metamorphic Approach to Single Quantum Dot Emission at 1.55 µm on GaAs Substrate*, Journal of Applied Physics **103**, 103533 (2008).

[44] L. Seravalli, P. Frigeri, G. Trevisi, and S. Franchi, *1.59 Mm Room Temperature Emission from Metamorphic InAs/InGaAs Quantum Dots Grown on GaAs Substrates*, Applied Physics Letters **92**, 213104 (2008).

[45] B. Scaparra, A. Ajay, P. S Avdienko, Y. Xue, H. Ried, P. Kohl, B. Jonas, B. Costa, E. Sirotti, P. Schmiedeke et al., *Structural Properties of Graded $In_xGa_{1-x}As$ Metamorphic Buffer Layers for Quantum Dots Emitting in the Telecom Bands*, Mater. Quantum Technol. **3**, 035004 (2023).

[46] P. A. Wroński, P. Wyborski, A. Musiał, P. Podemski, G. Sęk, S. Höfling, and F. Jabeen, *Metamorphic Buffer Layer Platform for 1550 µm Single-Photon Sources Grown by MBE on (100) GaAs Substrate*, Materials **14**, 18 (2021).

[47] R. Sittig, C. Nawrath, S. Kolatschek, S. Bauer, R. Schaber, J. Huang, P. Vijayan, P. Pruy, S. L. Portalupi, M. Jetter, and P. Michler, *Thin-Film InGaAs Metamorphic Buffer for Telecom C-Band InAs Quantum Dots and Optical Resonators on GaAs Platform*, Nanophotonics **11**, 1109 (2022).

[48] N. A. Maleev, A. V. Sakharov, C. Moeller, I. L. Krestnikov, A. R. Kovsh, S. S. Mikhrin, A. E. Zhukov, V. M. Ustinov, W. Passenberg, E. Pawlowski et al., *1300 nm GaAs-Based Microcavity LED Incorporating InAs/GaInAs Quantum Dots*, Journal of Crystal Growth **227–228**, 1146 (2001).

[49] F. Olbrich, J. Hoeschele, M. Mueller, J. Kettler, S. L. Portalupi, M. Paul, M. Jetter, and P. Michler, *Polarization-Entangled Photons from an InGaAs-Based Quantum Dot Emitting in the Telecom C-Band*, Appl. Phys. Lett. **111**, 133106 (2017).

[50] K. D. Zeuner, K. D. Jöns, L. Schweickert, C. R. Hedlund, C. Nuñez Lobato, T. Lettner, K. Wang, S. Gyger, E. Schöll, S. Steinhauer et al., *On-Demand Generation of Entangled Photon Pairs in the Telecom C-Band with InAs Quantum Dots*, ACS Photonics **8**, 2337 (2021).

[51] C. Nawrath, F. Olbrich, M. Paul, S. L. Portalupi, M. Jetter, and P. Michler, *Coherence and Indistinguishability of Highly Pure Single Photons from Non-Resonantly and Resonantly Excited Telecom C-Band Quantum Dots*, Applied Physics Letters **115**, 023103 (2019).

[52] C. Nawrath, H. Vural, J. Fischer, R. Schaber, S. L. Portalupi, M. Jetter, and P. Michler, *Resonance Fluorescence of Single In(Ga)As Quantum Dots Emitting in the Telecom C-Band*, Applied Physics Letters **118**, 244002 (2021).

[53] C. Nawrath, R. Joos, S. Kolatschek, S. Bauer, P. Pruy, F. Hornung, J. Fischer, J. Huang, P. Vijayan, R. Sittig et al., *Bright Source of Purcell-Enhanced, Triggered, Single Photons in the Telecom C-Band*, Adv. Quantum Technol. 2300111, 1 (2023).

[54] K. D. Zeuner, M. Paul, T. Lettner, C. Reuterskiöld Hedlund, L. Schweickert, S. Steinhauer, L. Yang, J. Zichi, M. Hammar, K. D. Jöns, and V. Zwiller, *A Stable Wavelength-Tunable Triggered Source of Single Photons and Cascaded Photon Pairs at the Telecom C-Band*, Applied Physics Letters **112**, 173102 (2018).

[55] T. Lettner, S. Gyger, K. D. Zeuner, L. Schweickert, S. Steinhauer, C. Reuterskiöld Hedlund, Sandra Stroj, Armando Rastelli, Mattias Hammar, Rinaldo Trotta, Klaus et al., *Strain-Controlled Quantum Dot Fine Structure for Entangled Photon Generation at 1550 µm*, Nano Lett. **21**, 10501 (2021).

[56] Ł. Dusanowski, C. Nawrath, S. L. Portalupi, M. Jetter, T. Huber, S. Klembt, P. Michler, and S. Höfling, *Optical Charge Injection and Coherent Control of a Quantum-Dot Spin-Qubit Emitting at Telecom Wavelengths*, Nat Commun **13**, 1 (2022).

[57] N. N. Ledentsov, A. R. Kovsh, A. E. Zhukov, N. A. Maleev, S. S. Mikhrin, A. P. Vasil'ev, E. S. Semenova, M. V. Maximov, Y. M. Shemyakov, N. V. Kryzhanovskaya et al., *High Performance Quantum Dot Lasers on GaAs Substrates Operating in 1.5 µm Range*, Electronics Letters **39**, 1126 (2003).

[58] L. Seravalli, G. Trevisi, G. Muñoz-Matutano, D. Rivas, J. Martinez-Pastor, and P. Frigeri, *Sub-Critical InAs Layers on Metamorphic InGaAs for Single Quantum Dot Emission at Telecom Wavelengths*, Crystal Research and Technology **49**, 540 (2014).

[59] P. Podemski, A. Maryński, P. Wyborski, A. Bercha, W. Trzeciakowski, and G. Sęk, *Single Dot Photoluminescence Excitation Spectroscopy in the Telecommunication Spectral Range*, Journal of Luminescence **212**, 300 (2019).

[60] K. Gawarecki, P. Machnikowski, and T. Kuhn, *Electron States in a Double Quantum Dot with Broken Axial Symmetry*, Phys. Rev. B **90**, 085437 (2014).

[61] T. B. Bahder, *Eight-Band Kp Model of Strained Zinc-Blende Crystals*, Phys. Rev. B **41**, 11992 (1990).

[62] A. Mielnik-Pyszczorski, K. Gawarecki, M. Gawełczyk, and P. Machnikowski, *Dominant Role of the Shear Strain Induced Admixture in Spin-Flip Processes in Self-Assembled Quantum Dots*, Phys. Rev. B **97**, 245313 (2018).

[63] M. Gawełczyk, M. Syperek, A. Maryński, P. Mrowiński, Ł. Dusanowski, K. Gawarecki, J. Misiewicz, A. Somers, J. P. Reithmaier, S. Höfling, and G. Sęk, *Exciton Lifetime and Emission Polarization Dispersion in Strongly In-Plane Asymmetric Nanostructures*, Phys. Rev. B **96**, 245425 (2017).

[64] A. Thränhardt, C. Ell, G. Khitrova, and H. M. Gibbs, *Relation between Dipole Moment and Radiative Lifetime in Interface Fluctuation Quantum Dots*, Phys. Rev. B **65**, 035327 (2002).

[65] J. Andrzejewski, G. Sęk, E. O'Reilly, A. Fiore, and J. Misiewicz, *Eight-Band K·p Calculations of the Composition Contrast Effect on the Linear Polarization Properties of Columnar Quantum Dots*, Journal of Applied Physics **107**, 073509 (2010).

[66] J. M. Ulloa, C. Çelebi, P. M. Koenraad, A. Simon, E. Gapihan, A. Letoublon, N. Bertru, I. Drouzas, D. J. Mowbray, M. J. Steer, and M. Hopkinson, *Atomic Scale Study of the Impact of the Strain and Composition of the Capping Layer on the Formation of InAs Quantum Dots*, Journal of Applied Physics **101**, 081707 (2007).

[67] V. M. Ustinov, N. A. Maleev, A. E. Zhukov, A. R. Kovsh, A. Yu. Egorov, A. V. Lunev, B. V. Volovik, I. L. Krestnikov, Yu. G. Musikhin, N. A. Bert et al., *InAs/InGaAs Quantum Dot Structures on GaAs Substrates Emitting at 1.3 µm*, Appl. Phys. Lett. **74**, 2815 (1999).

[68] A. Fiore, U. Oesterle, R. P. Stanley, R. Houdré, F. Lelarge, M. Ilegems, P. Borri, W. Langbein, D. Birkedal, J. M. Hvam et al., *Structural and Electrooptical Characteristics of Quantum Dots Emitting at 1.3 µm on Gallium Arsenide*, IEEE J. Quantum Electron. **37**, 1050 (2001).

[69] L. Seravalli, P. Frigeri, L. Nasi, G. Trevisi, and C. Bocchi, *Metamorphic Quantum Dots: Quite Different Nanostructures*, J. Appl. Phys. **108**, 064324 (2010).

[70] E. F. Schubert, E. O. Göbel, Y. Horikoshi, K. Ploog, and H. J. Queisser, *Alloy Broadening in Photoluminescence Spectra of $Al_xGa_{1-x}As$*, Phys. Rev. B **30**, 813 (1984).

[71] Ł. Dusanowski, A. Musiał, A. Maryński, P. Mrowiński, J. Andrzejewski, P. Machnikowski, J. Misiewicz, A. Somers, S. Höfling, J. P. Reithmaier, and G. Sęk, *Phonon-Assisted Radiative Recombination of Excitons Confined in Strongly Anisotropic Nanostructures*, Phys. Rev. B **90**, 125424 (2014).

[72] A. Musiał, G. Sęk, P. Podemski, M. Syperek, J. Misiewicz, A. Löffler, S. Höfling, and A. Forchel, *Excitonic Complexes in InGaAs/GaAs Quantum Dash Structures*, J. Phys.: Conf. Ser. **245**, 012054 (2010).

[73] Ł. Dusanowski, M. Syperek, A. Maryński, L. H. Li, J. Misiewicz, S. Höfling, M. Kamp, A. Fiore, and G. Sęk, *Single Photon Emission up to Liquid Nitrogen Temperature from Charged Excitons Confined in GaAs-Based Epitaxial Nanostructures*, Applied Physics Letters **106**, 233107 (2015).

[74] M. Abbarchi, F. Troiani, C. Mastrandrea, G. Goldoni, T. Kuroda, T. Mano, K. Sakoda, N. Koguchi, S. Sanguinetti, A. Vinattieri, M. Gurioli, *Spectral diffusion and line broadening in single selfassembled GaAs/AlGaAs quantum dot photoluminescence*, Applied Physics Letters **93**, 162101 (2008).

[75] M. Bayer, G. Ortner, O. Stern, A. Kuther, A. A. Gorbunov, A. Forchel, P. Hawrylak, S. Fafard, K. Hinzer, T. L. Reinecke et al., *Fine Structure of Neutral and Charged Excitons in Self-Assembled In(Ga)As/(Al)GaAs Quantum Dots*, Phys. Rev. B **65**, 195315 (2002).

[76] C. Zinoni, B. Alloing, C. Monat, V. Zwiller, L. H. Li, A. Fiore, L. Lunghi, A. Gerardino, H. de Riedmatten, H. Zbinden, and N. Gisin, *Time-Resolved and Antibunching Experiments on Single Quantum Dots at 1300 nm*, Appl. Phys. Lett. **88**, 86 (2006).

[77] M. Gschrey, F. Gericke, A. Schüßler, R. Schmidt, J. H. Schulze, T. Heindel, S. Rodt, A. Strittmatter, and S. Reitzenstein, *In Situ Electron-Beam Lithography of Deterministic Single-Quantum-Dot Mesa-Structures Using Low-Temperature Cathodoluminescence Spectroscopy*, Appl. Phys. Lett. **102**, (2013).

[78] S. V. Sorokin, G. V. Klimko, I. V. Sedova, A. A. Sitnikova, D. A. Kirilenko, M. V. Baidakova, M. A. Yagovkina, T. A. Komissarova, K. G. Belyaev, and S. V. Ivanov, *Peculiarities of Strain Relaxation in Linearly Graded $In_xGa_{1-x}As/GaAs(001)$ Metamorphic Buffer Layers Grown by Molecular Beam Epitaxy*, J. Cryst. Growth **455**, 83 (2016).

[79] C. Tonin, R. Hostein, V. Voliotis, R. Grousson, A. Lemaitre, and A. Martinez, *Polarization Properties of Excitonic Qubits in Single Self-Assembled Quantum Dots*, Phys. Rev. B **85**, 155303 (2012).

[80] J. P. Reithmaier, A. Somers, S. Deubert, R. Schwertberger, W. Kaiser, A. Forchel, M. Calligaro, P. Resneau, O. Parillaud, S. Bansropun, et al., *InP Based Lasers and Optical Amplifiers with Wire-/Dot-like Active Regions*, J. Phys. D: Appl. Phys. **38**, 2088 (2005).

[81] A. Löffler, J.-P. Reithmaier, A. Forchel, A. Sauerwald, D. Peskes, T. Kümmell, and G. Bacher, *Influence of the Strain on the Formation of GaInAs/GaAs Quantum Structures*, Journal of Crystal Growth **286**, 6 (2006).

[82] A. Musiał, P. Podemski, G. Sęk, P. Kaczmarkiewicz, J. Andrzejewski, P. Machnikowski, J. Misiewicz, S. Hein, A. Somers, S. Höfling et al., *Height-Driven Linear Polarization of the Surface Emission from Quantum Dashes*, Semicond. Sci. Technol. **27**, 105022 (2012).

[83] Y. Léger, L. Besombes, L. Maingault, and H. Mariette, *Valence-Band Mixing in Neutral, Charged, and Mn-Doped Self-Assembled Quantum Dots*, Phys. Rev. B **76**, 045331 (2007).

[84] A. V. Koudinov, I. A. Akimov, Yu. G. Kusrayev, and F. Henneberger, *Optical and Magnetic Anisotropies of the Hole States in Stranski-Krastanov Quantum Dots*, Phys. Rev. B **70**, 241305 (2004).

[85] G. Sęk, A. Musiał, P. Podemski, and J. Misiewicz, *On the Applicability of a Few Level Rate Equation Model to the Determination of Exciton versus Biexciton Kinetics in Quasi-Zero-Dimensional Structures*, Journal of Applied Physics **108**, 033507 (2010).

[86] Ł. Dusanowski, M. Syperek, W. Rudno-Rudziński, P. Mrowiński, G. Sęk, J. Misiewicz, A. Somers, J. P. Reithmaier, S. Höfling, and A. Forchel, *Exciton and Biexciton Dynamics in Single Self-Assembled InAs/InGaAlAs/InP Quantum Dash Emitting near 1.55 μm*, Appl. Phys. Lett. **103**, 253113 (2013).

[87] S. Rodt, R. Seguin, A. Schliwa, F. Guffarth, K. Pötschke, U. W. Pohl, and D. Bimberg, *Size-Dependent Binding Energies and Fine-Structure Splitting of Excitonic Complexes in Single InAs/GaAs Quantum Dots*, Journal of Luminescence **122–123**, 735 (2007).

[88] G. Sęk, P. Podemski, J. Misiewicz, L. H. Li, A. Fiore, and G. Patriarche, *Photoluminescence from a Single InGaAs Epitaxial Quantum Rod*, Appl. Phys. Lett. **92**, 021901 (2008).

[89] L. Seravalli, G. Trevisi, P. Frigeri, D. Rivas, G. Muñoz-Matutano, I. Suárez, B. Alén, J. Canet-Ferrer, and J. P. Martínez-Pastor, *Single Quantum Dot Emission at Telecom Wavelengths from Metamorphic InAs/InGaAs Nanostructures Grown on GaAs Substrates*, Applied Physics Letters **98**, 173112 (2011).

[90] K. Takemoto, M. Takatsu, S. Hirose, N. Yokoyama, Y. Sakuma, T. Usuki, T. Miyazawa, and Y. Arakawa, *An Optical Horn Structure for Single-Photon Source Using Quantum Dots at Telecommunication Wavelength*, Journal of Applied Physics **101**, 081720 (2007).

[91] A. Musiał, P. Holewa, P. Wyborski, M. Syperek, A. Kors, J. P. Reithmaier, G. Sęk, and M. Benyoucef, *High-Purity Triggered Single-Photon Emission from Symmetric Single InAs/InP Quantum Dots around the Telecom C-Band Window*, Advanced Quantum Technologies **3**, 1900082 (2020).

[92] N. Ha, T. Mano, S. Dubos, T. Kuroda, Y. Sakuma, and K. Sakoda, *Single Photon Emission from Droplet Epitaxial Quantum Dots in the Standard Telecom Window around a Wavelength of 1.55 μm*, Appl. Phys. Express **13**, 025002 (2020).

[93] P. A. Dalgarno, J. McFarlane, D. Brunner, R. W. Lambert, B. D. Gerardot, R. J. Warburton, K. Karrai, A. Badolato, and P. M. Petroff, *Hole Recapture Limited Single Photon Generation from a Single N-Type Charge-Tunable Quantum Dot*, Applied Physics Letters **92**, 193103 (2008).

[94] S. Fischbach, A. Schlehahn, A. Thoma, N. Srocka, T. Gissibl, S. Ristok, S. Thiele, A. Kaganskiy, A. Strittmatter, T. Heindel et al., *Single Quantum Dot with Microlens and 3D-Printed Micro-Objective as Integrated Bright Single-Photon Source*, ACS Photonics **4**, 1327 (2017).

[95] T. Miyazawa, K. Takemoto, Y. Nambu, S. Miki, T. Yamashita, H. Terai, M. Fujiwara, M. Sasaki, Y. Sakuma, M. Takatsu et al., *Single-Photon Emission at 1.5 μm from an InAs/InP Quantum Dot with Highly Suppressed Multi-Photon Emission Probabilities*, Appl. Phys. Lett. **109**, 132106 (2016).